\documentclass{raa}           
\usepackage{graphicx,times,subfigure}
\usepackage{natbib}
\usepackage{amssymb,amsmath}
\bibpunct{(}{)}{;}{a}{}{,}

\usepackage[a4paper=true,dvipdfm=true,pagebackref=true]{hyperref}
\hypersetup{pdftitle = The title of my PDF, pdfauthor = My name, pdfsubject= The subject, pdfkeywords = keyword1 keyword2 keyword3} 
\hypersetup{colorlinks = true, linkcolor = green, anchorcolor = red, citecolor = blue, filecolor = red, pagecolor = red, urlcolor = red}

\begin{document}

   \title{Near-Infrared Monitoring and Modeling of V1647 Ori in its On-going 2008-2012 Outburst Phase}

 \volnopage{ {\bf 2012} Vol.\ {\bf X} No. {\bf XX}, 000--000}
   \setcounter{page}{1}

   \author{V. Venkata Raman\inst{1}, B.G. Anandarao\inst{1}, P. Janardhan\inst{1},
   R. Pandey \inst{2}}

   \institute{ Physical Research Laboratory (PRL), Ahmedabad - 380009, India;
   {\it vvenkat@prl.res.in}\\
        \and
             Physics Department, M.L.S. University, Udaipur, India\\
	\vs \no
   {\small Received ...; accepted ...}
}

\abstract{We present results of Mt Abu {\it JHK} photometric and HI Brackett $\gamma$ line 
monitoring of the eruptive YSO V1647 Orionis (McNeil's Object) during the 
on-going outburst phase in 2008-2012. We discuss {\it JHK} color patterns and 
extinction during the outburst and compare them with those in the previous outburst 
phase in 2004-2005 and in the intervening quiescent period of about 2 years. Commencing from  
early 2012, the object shows a slow fading out in all the bands.   
We report brightness variations in the nearby Herbig-Haro object HH22 possibly 
associated with those in V1647 Ori. We also present modeling of the spectral energy 
distributions of V1647 Ori during both its recent outburst and its quiescent phase. 
The physical parameters of the protostar and its circumstellar environment 
obtained from the modeling indicate marked differences between the two phases. 
}

   \authorrunning{V. Venkata Raman et al. }            
   \titlerunning{Near-infrared monitoring and modeling of V1647 Ori}  
   \maketitle

%
\section{Introduction}           
\label{sect:intro}

It is believed that most of the low-mass pre-main-sequence (PMS) stars undergo a recurring 
active stage during which they show enhanced brightness or an `outburst' lasting for a few years to 
a decade or longer (e.g., \cite{Stahler+Palla+2004} and references therein). Among the various mechanisms 
proposed (see \cite{Hartmann+Kenyon+1996}), by which the rate of accreted mass increases substantially 
over a period of time causing the outburst, are the thermal instability 
in accretion disks (\cite{Bell+etal+1995}), evolution from envelope accretion to disk 
magnetospheric accretion (\cite{Hartmann+Kenyon+1985}), 
and gravitational or tidal triggering by the passage of a putative binary companion 
(\cite{Bonnell+Bastien+1992}). Depending upon the duration of the outburst, PMS stars  
are sub-classified as FUors (prototype being FU Orionis) that last for over 
a decade or longer and EXors (prototype EX Lupis) that occur for shorter periods
of about 
two years (\cite{Herbig+1977}). V1647 Ori (IRAS 05436-0007), 
first discovered by \cite{Mcneil+2004} and since called  
McNeil's object, showed an outburst akin to EXors, yet it seemed to be distinctive 
(e.g., \cite{Ojha+etal+2006}; \cite{Aspin+etal+2008}; 
 \cite{Aspin+Reipurth+2009} and the references therein). 
While its outburst duration was similar to EXors, its spectral characteristics seemed to be so
different from EXors that it may as well be a new class of object in itself (\cite{Kun+2008}). 
However, in recent times the star showed a second outburst (currently on-going) starting from 
mid-2008 (for a comprehensive study, see \cite{Aspin+2011} 
and references therein). Reporting post-2008 outburst behavior,  \cite{Aspin+2011} concluded 
that the object remained ``in an elevated photometric state" till early 2011.
It was also concluded that McNeil's nebular morphology remained unchanged in its two recent outbursts.
Furthermore a large discrepancy in the accretion rates derived from H$\alpha$ and Br$\gamma$ emission
line fluxes was reported. Using high resolution spectroscopy,  \cite{Brittain+etal+2010} showed that the accretion rates
derived from Br$\gamma$ emission were similar during the two outbursts (though varying) and a factor of $\sim$
16 higher than the smallest accretion rate during the quiescent phase. 
These studies showed that the current (on-going) outburst was basically different from the 
earlier one and this deserves further investigation.  
Clearly, therefore, continued monitoring of the object is necessary to detect 
possible changes in its behavior post-2011.    
 
In this work, we present a substantial volume of {\it JHK} photometric and some {\it K} band spectroscopic observations 
made from Mt Abu during 2008-2012. We discuss their possible implications on the nature of V1647 Ori, in comparison with
the previous outburst and quiescent phases. 
Further, we attempt to model the Spectral Energy Distributions (SED) 
using an on-line modeling tool developed by \cite{Robitaille+etal+2007} and compare the physical 
parameters including accretion rates of both disk and envelope of V1647 Ori
during the two epochs (viz. quiescence and outburst).  
Section 2 gives the details of our observations and 
Section 3 presents the results and discussion, including SED modeling results. 
Section 4 gives important conclusions.

\section{Near-Infrared Photometric and Spectroscopic Observations}
\label{sect:Obs}

{\it JHK} photometric observations were made using the 
Near Infrared Camera \& Multi-object Spectrograph (NICMOS) (256x256 HgCdTe array) 
and the Near Infrared Camera \& Spectrograph (NICS) (HAWAII-1 1kx1k HgCdTe array) 
at the Cassegrain focus of the 1.2m infrared telescope
situated at PRL's Mt Abu Observatory. 
The data consists of more than 40 sets of {\it JHK} observations during the period 2004-2012, a majority of which were during
the period 2008-2012. 
A part of the photometric data obtained between 2004-2005 had appeared in \cite{Ojha+etal+2006}. 
Single frame integration times were 40-60s/20-30s for {\it J}, 20-40s/10-20s for {\it H} and 2s/10-15s for
{\it K} band for the NICMOS/NICS cameras. Several such frames were taken amounting to total 
integration times of 120-720s for NICMOS and 120-400s for NICS.
A sufficient number of dithered frames were obtained for effective background subtraction and flat fielding. 
The seeing during the observations was typically 1.7-2.5 arcsec. 
Photometric flux calibration was done by observing a standard star in the AS 13 region (\cite{Hunt+etal+1998}).  
Data reduction was done using the Image reduction and Analysis Facility (IRAF) software.
The dark-subtracted and background-subtracted images were co-added to obtain the final image in each band.
The photometric magnitudes of V1647 Ori were then found using the task APPHOT in {\sc IRAF}. 
The magnitude of extended sources such as HH22 was also estimated using a larger sampling aperture
(usually 4 times the FWHM of a star image).
The integrated magnitude of the nebula surrounding V1647 Ori was estimated from the star-subtracted 
images using a 40 arcsec aperture.
{\it K} band spectroscopic observations were made using the NICMOS array at a spectral solution of $\sim$ 1000. 
The integration time for the spectra was 60s per frame. 
Spectral reduction was done using standard spectroscopic tasks in {\sc IRAF}.
For sky background subtraction, a set of 
at least two spectra were taken with the object dithered to two different
positions along the slit. The sky-subtracted spectra were then co-added resulting in a 
total exposure time of 480s. The atmospheric OH vibration-rotation lines were
used for wavelength calibration.
The spectra of V1647 Ori was then ratioed with that of a spectroscopic standard star of AOV type, 
observed at a similar airmass, to remove the telluric absorption features. Prior to ratioing, the HI absorption lines from the standard 
star spectra were removed by interpolation. The ratioed spectra were then multiplied by a blackbody curve at the 
effective temperature of the standard star to yield the final spectra. The observed K band photometric flux of V1647 Ori
was used for flux calibration of the spectra.

\section{Results and Discussion}
\subsection{Photometry}

Table 1 gives the {\it JHK} magnitudes and colors for all the dates of our observations.
The photometric errors are typically 0.02 to 0.05 magnitudes.
For the on-going 2008 outburst period 
(observations from Nov 2008 till Nov 2012), the average magnitudes with standard deviation ($\sigma$) for 29 data sets are: 10.85 (0.17); 9.02 (0.18); 
7.60 (0.24) for the {\it J}, {\it H}, and {\it K} bands respectively. Occasional deviations of more than 3$\sigma$ due to variability of object were noticed in all the three bands.
For the 2004 outburst period (observations from Mar 2004 till Dec 2004) 
the averages for 7 data sets are: 11.01 (0.16); 9.06 (0.17); 7.73 (0.15) for the {\it J}, {\it H}, and {\it K} bands respectively; the 
deviations were within 2$\sigma$.  
For the quiescent phase during 2006-2007 (observations from Dec 2005 till Dec 2006):
the averages for 8 data sets are : 14.2 (0.18); 11.61 (0.14); 9.89 (0.18) for the {\it J}, {\it H}, and {\it K}
 bands respectively; the occasional deviations seen were within 2$\sigma$.
Thus the on-going outburst shows higher amplitude fluctuations in brightness especially 
in {\it K} band (more than 0.5 mag) than the 2004-2005 outburst. The light curves in the {\it J}, {\it H} 
and {\it K} bands are shown in Fig 1 covering the period between 2004-2012. For comparison, a few 
data points from other published work are also shown (from \cite{Peregrine+etal+2004},  
\cite{Ojha+etal+2006}, \cite{Acosta-Pulido+etal+2007}, and \cite{Aspin+2011}).  
It may be mentioned here that \cite{Acosta-Pulido+etal+2007} reported a 56 day periodic component 
in their optical light curves in the visible bands. Our {\it JHK} data do not clearly show this component. 

\begin{figure}
\centering
\includegraphics[width=0.85\textwidth, angle=-90.0]{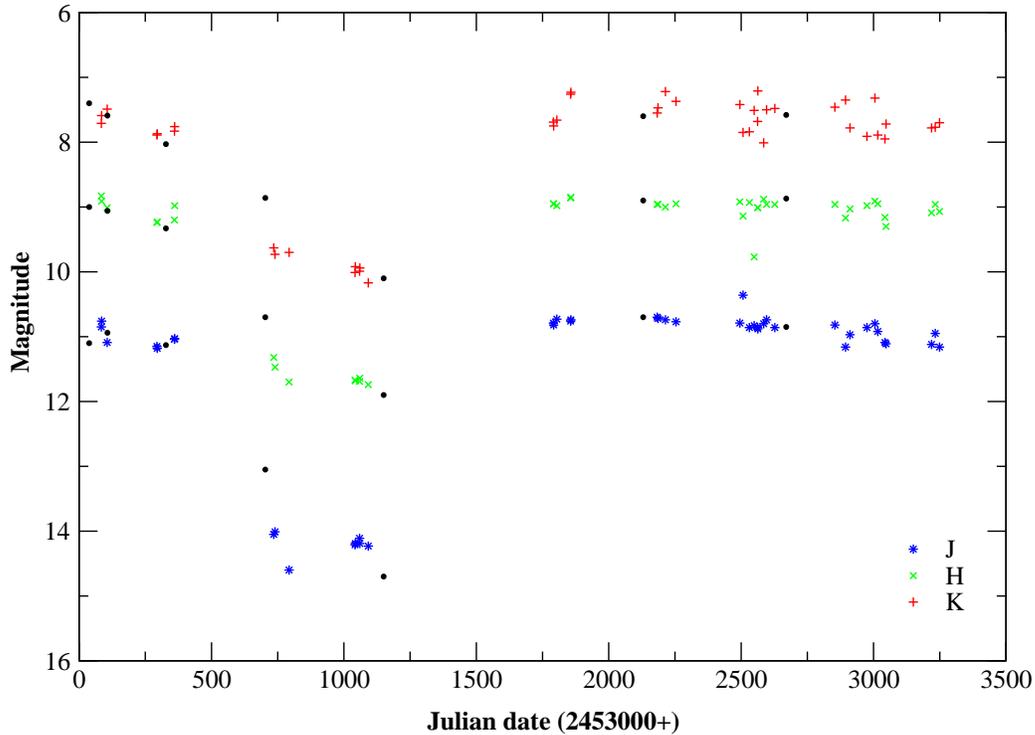}
\caption{V1647 Ori: Light Curves in the {\it J}, {\it H}, {\it K} bands from Mt Abu observations
 during 2004-2012 covering the two outbursts with the $\sim$ 2-yr quiescence in 
 between. The blue asterisks represent the {\it J} band, green crosses represent the {\it H} band and
red pluses represent the {\it K} band.
Observations made elsewhere are shown in black filled circles (see text for references).} 
\label{fig1}
\end{figure}

{\it{Trend of decline of Outburst phase since early 2012:}} 
In comparison with its behavior till 2011 (e.g., \cite{Venkat+Anandarao+2011}), the object seems to show a
steady decline in the brightness beginning from Feb-Mar 2012 (cf. \cite{Semkov+Peneva+2012}, 
\& \cite{Ninan+etal+2012}) with an approximate rate of 0.16, 0.06, 0.18 mag per year in the {\it JHK} bands respectively (see Fig 1).
It is necessary to confirm this declining trend by continued monitoring. The slow decline
compared to the one noticed in the 2004 outburst is reminiscent of a typical FUors light curve. But it is 
premature to conclude that it is a FUor; as there have been spectroscopic indications to the 
contrary. In fact it could be a class in itself which shows both FUors and EXors characteristic features
(see \cite{Kun+2008}).

We have computed the visual extinction A$_{V}$ from the colors [J-H] and [H-K] 
for both the epochs, using the formula for T Tauri stars derived from 
\cite{Meyer+etal+1997} and the extiction model of \cite{Bessell+Brett+1988},

\begin{equation} 
A_{V} = 13.83 \times [J-H] - 8.02 \times [H-K] - 7.19 
\end{equation}
The computed A$_{V}$
values are listed in Table 1. For comparison, the 
 A$_{V}$ corresponding to 2MASS 
epoch (quiescent on 1998 October 7) is 13.3. The time evolution of A$_{V}$ is shown in Fig 2. 
We find that there is $\sim$ 6 mag difference in A$_{V}$ between 
the two phases - the outburst phase having lower extinction. This can be attributed to the 
excess mass accreted by the envelope (from the environment) during the quiescent phase in comparison with 
the outburst phase during which the disk accretion (from the envelope) is expected to dominate
(e.g.,  \cite{Aspin+etal+2008} and \cite{Aspin+2011}).  

\begin{figure*}
\centering
\includegraphics[width=0.85\textwidth, angle=-90.0]{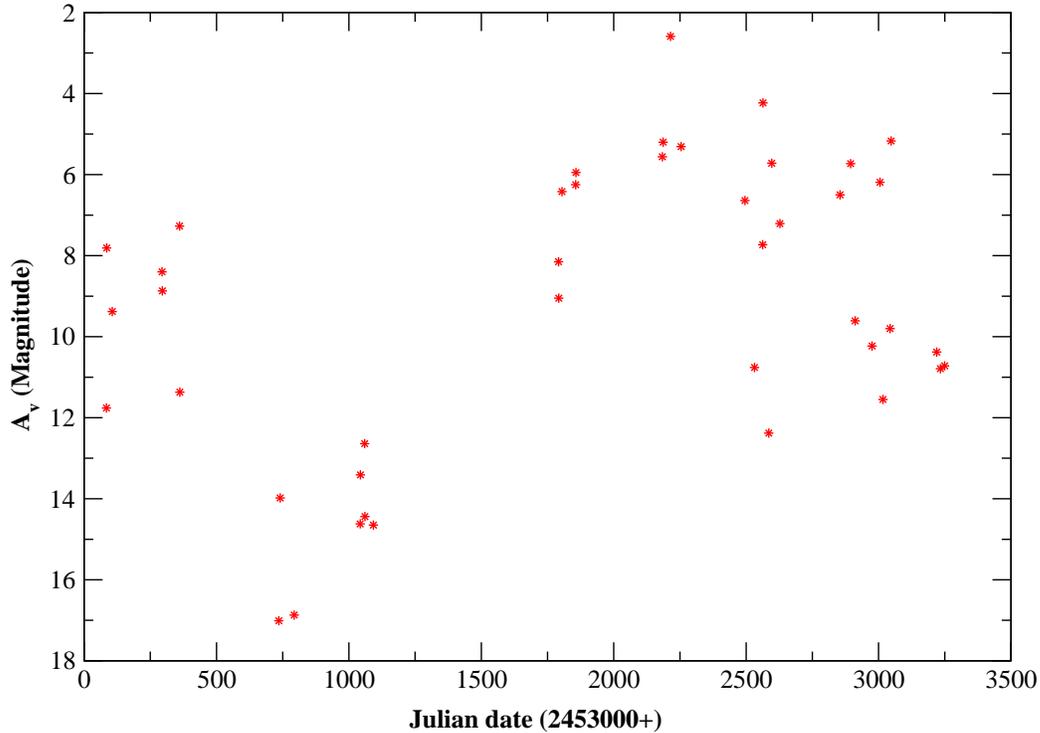}
\caption{V1647 Ori: Time variation of A$_{V}$ from Mt Abu observations during 2004-2012
comprising of two outbursts flanking the quiescent epoch (of larger extinction).} 
\label{fig3}
\end{figure*}

A {\it JHK} color-color diagram constructed from 
Mt Abu data during 2004-2012 is shown in Fig 3. 
The A$_{V}$ values computed using Eqn.(1) are
in reasonable agreement (within 1 - 2 mag) with those obtained by de-reddening 
the JHK colors to the T Tauri locus in the color-color diagram.
The color-color diagram demonstrates
the variable nature of the source in the on-going 2008 outburst phase in comparison with the 
2004 outburst. The region occupied by the latest outburst (asterisks in Fig 3) extends 
horizontally (i.e., with [J-H] $\sim$ 1.7-2.0) beyond the T Tauri region; quite in contrast with the 
2004 outburst phase (open circles). 
This tendency during the on-going outburst indicates the 
presence of cold dust in the envelope/disk of the star compared to the 2004 outburst. 
Also, we found at least two occasions (in 2010 November 7 \& 2010 December 19) on which the 
colors showed extreme values - as indicated by the magenta asterisks in Fig 3 
at around [J-H] $\sim$ 1.0-1.2, with [H-K] at 1.3 and 2.3. On another occasion (2011 January 24) the colors
indicate a position to the left of the T Tauri region. The A$_{V}$ could
not be calculated for these cases falling well beyond the T Tauri regime
(shown as dashes in Table 1). 
We suspect that such extreme fluctuations may be of 
short duration (a few days) and attributable to circumstellar dust temperature variation.

\begin{figure*}
\centering
\includegraphics[width=0.85\textwidth, angle=-90.0]{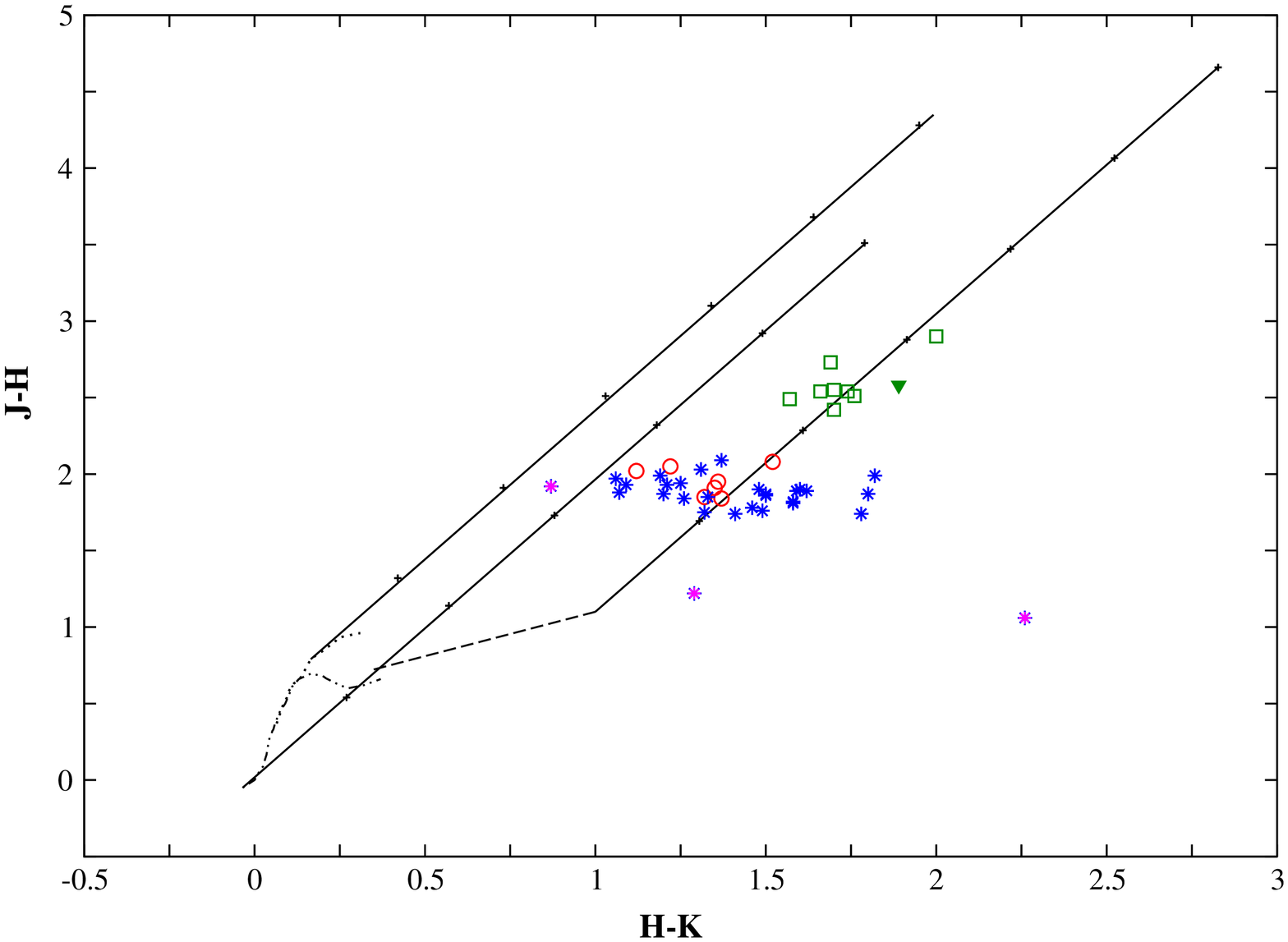}
\caption{V1647 Ori: Color-color diagram from Mt Abu observations. 
The 2MASS value is marked as inverted green triangle (corresponding to 1998 October 7). 
The blue asterisks correspond to the on-going outburst period 2008-2012; while the 
red circles correspond to the outburst period of 2004-2005. The green squares  
are for the quiescent epoch of 2006-2007. The three "stray" magenta asterisks, 
two appearing around J-H $\sim$ 1.0 and the one appearing around H-K $\sim$ 0.8 correspond 
to extremities in photometric 
magnitudes observed in our monitoring. See text for discussion. The dot-dashed 
curve represents the unreddened main-sequence stars; the dotted   
curve shows the unreddened giant stars. The black slanting parallel lines 
indicate the extinction vectors for A$_{V}$ = 30 mag (\cite{Bessell+Brett+1988} 
extinction law was adopted). The pluses along the extinction vector
corresponds to an extinction interval of A$_{V}$ = 5 mag. The T Tauri stars fall on the black dashed straight line, 
adopted from \cite{Meyer+etal+1997}).} 
\label{fig3}
\end{figure*}   

\begin{table}
\centering
\caption{V1647 Ori: JHK Photometric Magnitudes and Colors and A$_{V}$ (in mag) during 2004-2012} 
\label{tab2}
\begin{tabular}{lccccccr}
\hline
Date       &     JD     &   J    &   H    &    K   &   J-H  &   H-K  &  A$_{V}$ \\                                     \\
\hline
2004Mar20  &  2453084.5 &  10.85 &   8.83 &   7.71 &   2.02 &   1.12 &    11.76  \\
2004Mar21  &  2453085.5 &  10.76 &   8.91 &   7.59 &   1.85 &   1.32 &     7.81  \\
2004Apr11  &  2453106.5 &  11.09 &   9.01 &   7.49 &   2.08 &   1.52 &     9.38  \\
2004Oct16  &  2453294.5 &  11.15 &   9.24 &   7.89 &   1.91 &   1.35 &     8.40  \\
2004Oct17  &  2453295.5 &  11.18 &   9.23 &   7.87 &   1.95 &   1.36 &     8.87  \\
2004Dec21  &  2453360.5 &  11.04 &   9.20 &   7.83 &   1.84 &   1.37 &     7.27  \\
2004Dec22  &  2453361.5 &  11.03 &   8.98 &   7.76 &   2.05 &   1.22 &    11.37   \\
2005Dec31  &  2453735.5 &  14.05 &  11.32 &   9.63 &   2.73 &   1.69 &    17.01   \\
2006Jan05  &  2453740.5 &  14.01 &  11.47 &   9.73 &   2.54 &   1.74 &    13.98   \\
2006Feb27  &  2453793.5 &  14.60 &  11.70 &   9.70 &   2.90 &   2.00 &    16.87   \\
2006Nov03  &  2454042.5 &  14.21 &  11.67 &  10.01 &   2.54 &   1.66 &    14.62  \\
2006Nov04  &  2454043.5 &  14.19 &  11.68 &   9.92 &   2.51 &   1.76 &    13.41   \\
2006Nov20  &  2454059.5 &  14.11 &  11.69 &   9.99 &   2.42 &   1.70 &    12.64   \\
2006Nov21  &  2454060.5 &  14.19 &  11.64 &   9.94 &   2.55 &   1.70 &    14.44   \\
2006Dec23  &  2454092.5 &  14.23 &  11.74 &  10.17 &   2.49 &   1.57 &    14.65   \\
2008Nov21  &  2454791.5 &  10.79 &   8.95 &   7.69 &   1.84 &   1.26 &     8.15   \\
2008Nov22  &  2454792.5 &  10.82 &   8.95 &   7.75 &   1.87 &   1.20 &     9.05   \\
2008Dec04  &  2454804.5 &  10.73 &   8.98 &   7.66 &   1.75 &   1.32 &     6.42   \\
2009Jan25  &  2454856.5 &  10.76 &   8.86 &   7.26 &   1.90 &   1.60 &     6.25  \\
2009Jan26  &  2454857.5 &  10.74 &   8.85 &   7.23 &   1.89 &   1.62 &     5.95   \\
2009Dec18  &  2455183.5 &  10.70 &   8.96 &   7.55 &   1.74 &   1.41 &     5.56   \\
2009Dec21  &  2455186.5 &  10.72 &   8.96 &   7.47 &   1.76 &   1.49 &     5.20  \\
2010Jan18  &  2455214.5 &  10.74 &   9.00 &   7.22 &   1.74 &   1.78 &     ---   \\
2010Feb27  &  2455254.5 &  10.77 &   8.95 &   7.37 &   1.82 &   1.58 &     5.31   \\
2010Oct26  &  2455495.5 &  10.79 &   8.92 &   7.42 &   1.87 &   1.50 &     6.64   \\
2010Nov07  &  2455507.5 &  10.36 &   9.14 &   7.85 &   1.22 &   1.29 &      ---   \\
2010Dec01  &  2455531.5 &  10.86 &   8.93 &   7.84 &   1.93 &   1.09 &    10.76   \\
2010Dec19  &  2455549.5 &  10.83 &   9.77 &   7.51 &   1.06 &   2.26 &      ---   \\
2011Jan01  &  2455562.5 &  10.86 &   9.01 &   7.68 &   1.85 &   1.33 &     7.73   \\
2011Jan02  &  2455563.5 &  10.88 &   9.01 &   7.21 &   1.87 &   1.80 &     ---   \\
2011Jan24  &  2455585.5 &  10.80 &   8.88 &   8.01 &   1.92 &   0.87 &     ---   \\
2011Feb04  &  2455596.5 &  10.74 &   8.96 &   7.50 &   1.78 &   1.46 &     5.72   \\
2011Mar07  &  2455627.5 &  10.86 &   8.96 &   7.48 &   1.90 &   1.48 &     7.21  \\
2011Oct20  &  2455854.5 &  10.82 &   8.96 &   7.46 &   1.86 &   1.50 &     6.50   \\
2011Nov29  &  2455894.5 &  11.16 &   9.17 &   7.35 &   1.99 &   1.82 &     ---  \\
2011Dec16  &  2455911.5 &  10.97 &   9.03 &   7.78 &   1.94 &   1.25 &     9.61  \\
2012Feb18  &  2455975.5 &  10.86 &   8.98 &   7.91 &   1.88 &   1.07 &    10.23  \\
2012Mar19  &  2456005.5 &  10.80 &   8.91 &   7.32 &   1.89 &   1.59 &     6.19   \\
2012Mar30  &  2456016.5 &  10.92 &   8.95 &   7.89 &   1.97 &   1.06 &    11.55   \\
2012Apr26  &  2456043.5 &  11.09 &   9.16 &   7.95 &   1.93 &   1.21 &     9.80   \\
2012Apr30  &  2456047.5 &  11.11 &   9.30 &   7.72 &   1.81 &   1.58 &     5.17   \\
2012Oct19  &  2456219.5 &  11.12 &   9.09 &   7.78 &   2.03 &   1.31 &    10.38   \\
2012Nov02  &  2456233.5 &  10.95 &   8.96 &   7.77 &   1.99 &   1.19 &    10.79   \\
2012Nov18  &  2456249.5 &  11.16 &   9.07 &   7.70 &   2.09 &   1.37 &    10.72   \\
\hline 
\end{tabular}
\end{table}

\subsection{Spectroscopic Variations} As may be seen in Fig 4, {\it K} band spectra taken at different times during the outburst 
phase show variability in Hydrogen Br$\gamma$ indicating variable disk accretion rates. 
Br$\gamma$ line shows fluctuating trend that does not seem 
to be associated with photometric fluctuations 
(cf. H$\alpha$ line reported by  \cite{Walter+etal+2004} and \cite{Aspin+Reipurth+2009}). 
\cite{Quanz+etal+2007} reported molecular line variability in the mid-IR region during the 2004 outburst.
It may be noted that the width at zero-intensity and the ratio of
peak to continuum intensity in Mt. Abu profiles are in reasonable agreement 
with those of \cite{Brittain+etal+2010} for the year 2009, in spite of the large differences in the 
resolving power employed and S/N values between the two observations. 
The accretion luminosities and disk accretion rates derived from the dereddened 
Br $\gamma$ line fluxes from our spectra (following the procedure described by \cite{Muzerolle+etal+1998})
range from 20 - 60 L$_{\odot}$ and 1.0$\times10^{-6}$ to 3.0$\times10^{-6}$M$_{\odot}yr^{-1}$ respectively
which are comparable with those reported by \cite{Brittain+etal+2010}.

\begin{figure*}
\centering
\includegraphics[width=0.75\textwidth, angle=-90.0]{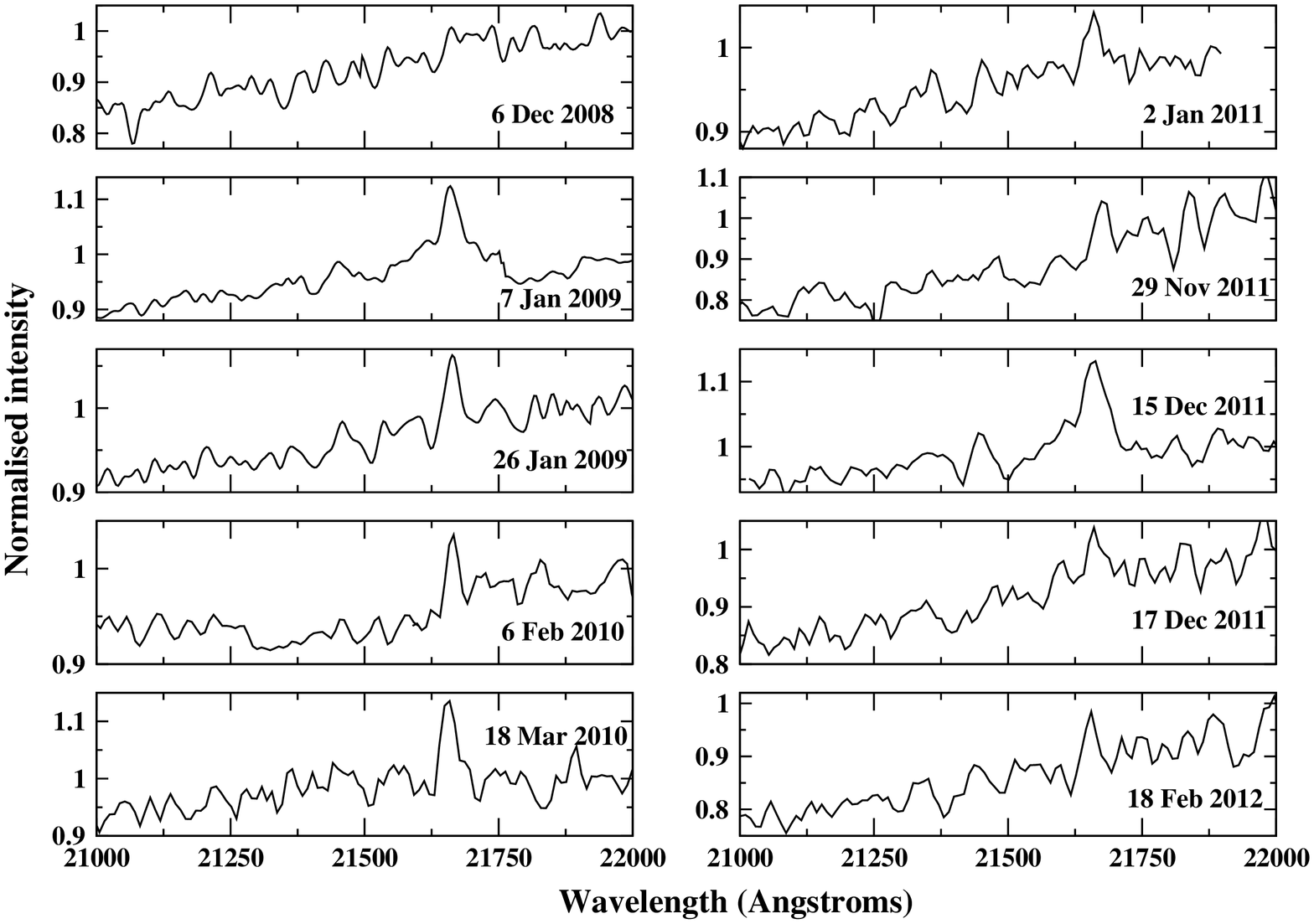}
\caption{V1647 Ori: Br$\gamma$ line (at 2.16$\mu$m) in the K band observed on a few 
occasions at Mt Abu. Variability is evident.} 
\label{fig3}
\end{figure*} 

\begin{figure*}
\centering
\includegraphics[width=0.90\textwidth]{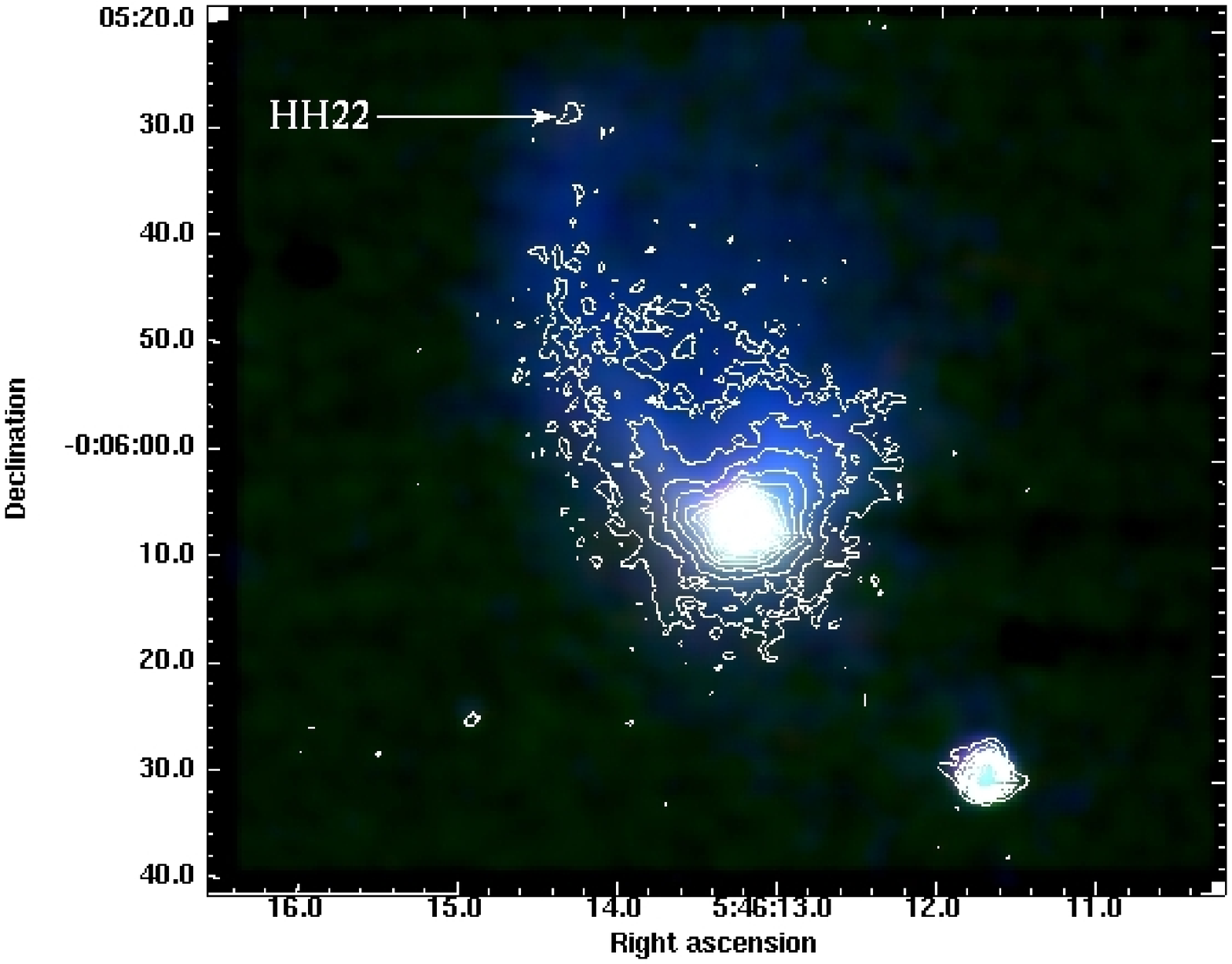}\\
\caption{Color composite image of V1647 Ori and the surrounding McNeil's nebula
in the {\it J}(blue), {\it H}(green) and {\it K}(red) bands taken at Mt Abu using NICS on 2011 February 4; 
HH22 is seen at nearly the north of V1647 Ori. 
The {\it H} Band contours (in white) are shown superposed on the image: the outermost contour is 
20 mag arcsec$^{-2}$ with each contour brightening by 2 mag arcsec$^{-2}$ inwards.}
\label{fig3}
\end{figure*}

\subsection{Variations in HH22 and the nebula around V1647 Ori} 
Fig 5 shows a {\it JHK} color composite image of V1647 Ori and its 
associated nebula taken from Mt Abu on 2011 February 04 using the NICS. Superposed on the image are 
the contours of iso-magnitudes in {\it H} band. The curved tail at the top left part of the nebular 
object is usually attributed to the on-going accretion
(cf. \cite{Reipurth+Aspin+2004}). HH22 (knot A in \cite{Eisloffel+Mundt+1997}), 
seen nearly to the north of V1647 Ori (see Fig 5), is a Herbig-Haro type object
(for mid-IR counterpart, see \cite{Muzerolle+etal+2005}). 
It happens to be present very close to V1647 Ori (30 arcsec of separation at an assumed 
distance of 0.40 kpc) and its originating source is not yet identified.
It could possibly be a reflection nebulosity powered by V1647 Ori (see, \cite{Briceno+etal+2004} 
and \cite{Aspin+etal+2008}). It may be expected
therefore that the outburst from V1647 Ori could cause the nebulosity to show a corresponding brightening.  
The light travel time from the YSO to HH22 is $\sim$ 70 days.
We analyzed the images in {\it JHK} bands obtained from Mt Abu to see if HH22 shows any variability in its 
brightness. Fig 6 shows integrated {\it K} magnitude of HH22 with time along with V1647 Ori magnitudes.  
The plot indicates a possible correspondence of HH22 (knot A) with the outburst activity; as well as 
indication of trigger from even short term fluctuations from V1647 Ori.
We also noticed fluctuations in the brightness of the nebula around V1647 Ori which typically follow
those of V1647 Ori itself. Similar fluctuations were also noticed in the J and H bands.

\begin{figure}
\centering
\includegraphics[width=0.85\textwidth, angle=-90.0]{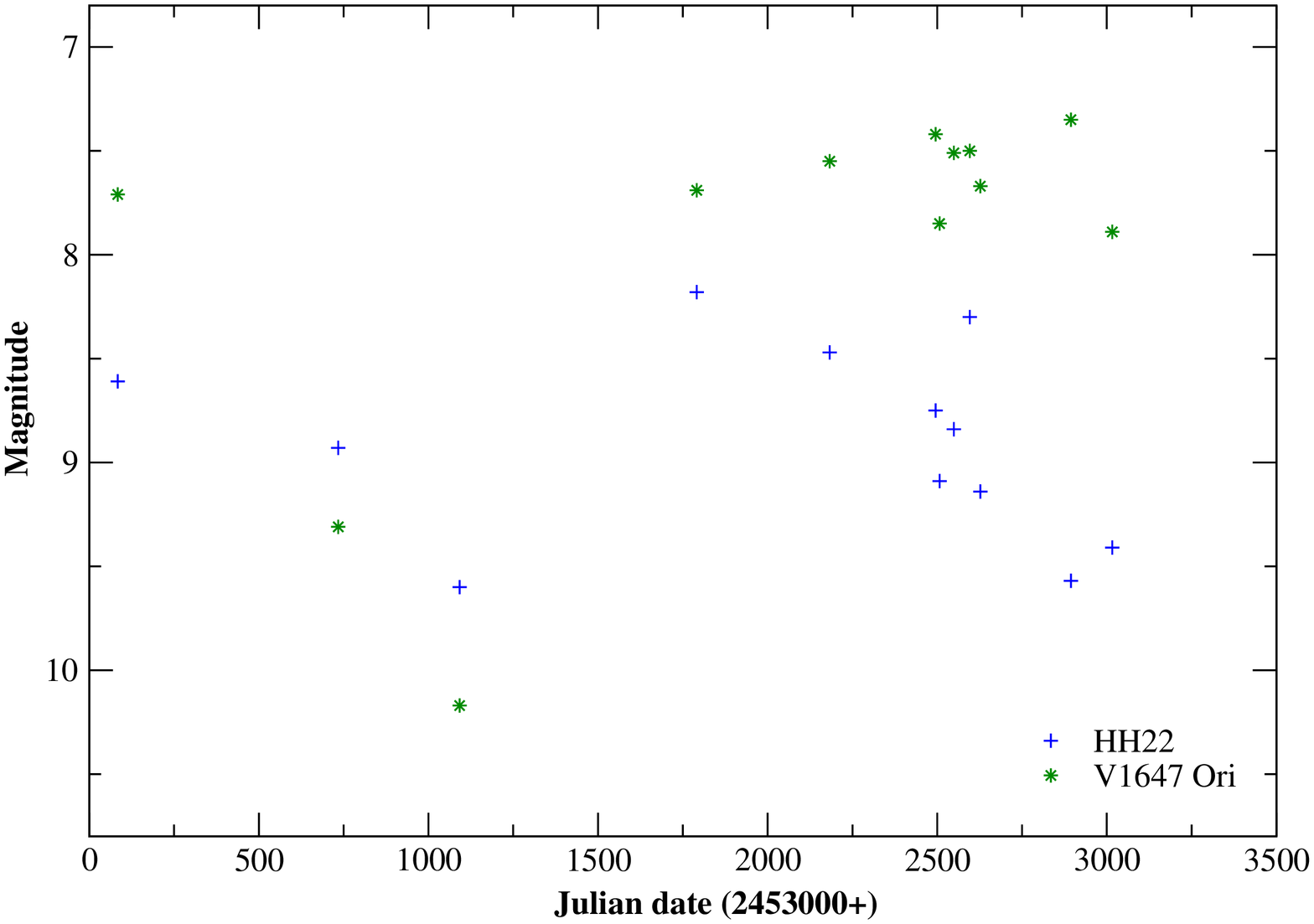}
\caption{Time variations of V1647 Ori in the {\it K} band seen against those in HH22.
The green asterisks correspond to V1647 Ori and blue pluses represent HH22. 
For HH22 the magnitudes are made brighter by subtracting 5 
to facilitate a closer comparison.} 
\label{fig3}
\end{figure}

\begin{figure}
	\begin{center}
	\subfigure[]{
	\includegraphics[width=0.7\textwidth,angle=-90.0]{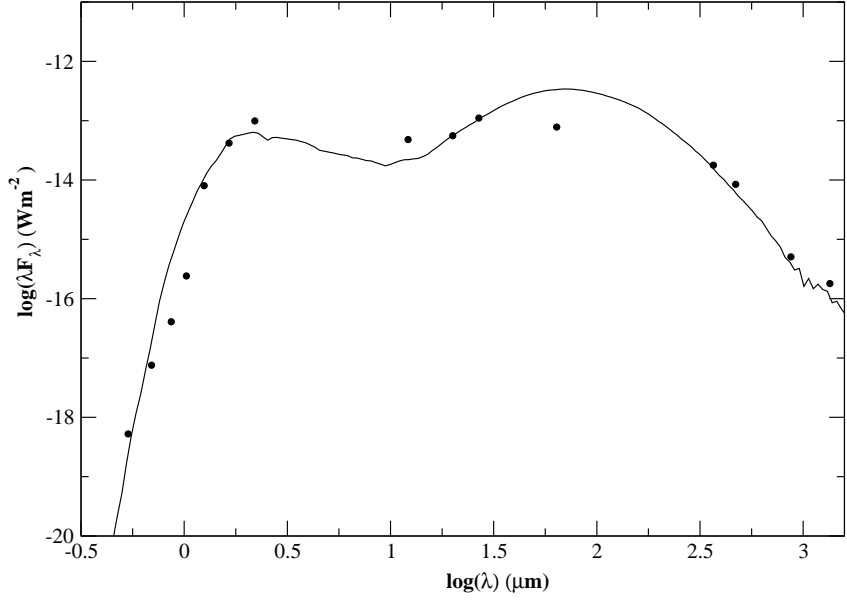}}\\
	\subfigure[]{
	\includegraphics[width=0.7\textwidth,angle=-90.0]{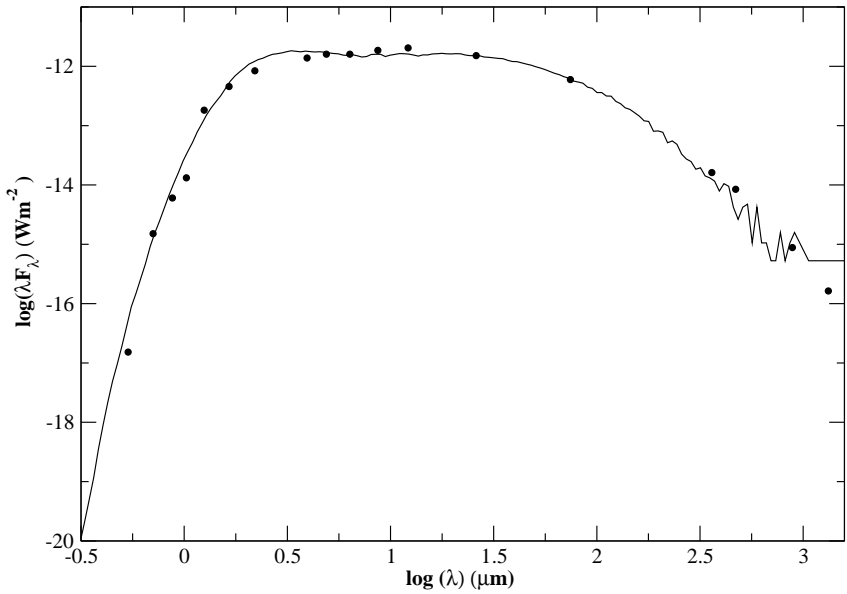}}\\
	\caption{SED models for V1647 Ori during (a) its quiescent phase (2006-2007) and (b) its 
	 outburst phase (2008-2012). The filled circles represent data points. The full dark curves show the best fits 
       (see text for details).}
	\end{center}
\label{fig1}
\end{figure}

\begin{table}
\centering
\caption{Model Parameters for V1647 Ori in its quiescent (2006-2007) and outburst (2008-2012) phases} 
\label{tab2}
\begin{tabular}{lcc}
\hline
Physical Parameter                          &  Quiescent Phase            &  Outburst Phase              \\
\hline                                                                                                                     
Stellar Age (yr)                            &  4.8$\pm$2.5$\times10^{4}$  &  6.3$\pm$3.1$\times10^{5}$      \\  
Star Mass (M$_{\odot}$)                     &  0.34$\pm$0.15               &  4.44$\pm$0.34                \\
Star Temperature (T (K))                    &  3360$\pm$170                &   7560$\pm$1590             \\
Star Radius (R$_{\odot}$)                   &  4.01$\pm$0.12               &  7.08$\pm$2.28                \\
Total Luminosity (L$_{\odot}$)              &  1.85$\pm$0.24                 &  156.5$\pm$28.0                 \\
Envelope Mass (M$_{\odot}$)                 &  1.51$\pm$0.02                &  1.40$\pm$3.79$\times10^{-4}$  \\
Envelope Accretion Rate (M$_{\odot}yr^{-1}$)   &  5.3$\pm$0.1$\times10^{-5}$  &  3.7$\pm$0.0$\times10^{-6}$  \\                
Disk Mass (M$_{\odot}$)                     &  6.8$\pm$8.00$\times10^{-5}$  &  6.9$\pm$1.8$\times10^{-2}$      \\
Disk Accretion Rate (M$_{\odot}yr^{-1}$)       &  5.5$\pm$3.3$\times10^{-9}$  &  2.2$\pm$1.6$\times10^{-7}$    \\
\hline 
\end{tabular}
\end{table}

\subsection{Modeling of Spectral Energy Distribution}

Modeling the SED of V1647 Ori was done for three sets of data taken during : (i) the outburst phase 2004-2005; 
(ii) the quiescent phase 2006-2007 and (iii) the second (on-going) outburst phase 2008-2012. 
For each of these three phases the mean values of Mt Abu {\it JHK} magnitudes were considered.
To the {\it JHK} data we added the visible, mid-infrared, far infrared and mm-wave data on quiescent and 
outburst epochs taken from \cite{Andrews+etal+2004} and \cite{Aspin+etal+2008}. 
The SEDs given in \cite{Aspin+etal+2008} for the outburst and quiescent phases show that the 
fluxes of the PMS star in sub-mm and mm regions did not change more than 10\% between the two 
phases. It is in the visible and infrared that a substantial change had taken place. 
The near-infrared region occurs right at the position of turn-over in the SED and hence is quite important.
It was assumed that the small photometric variations, if present in the mid-infrared and far-infrared, 
are not significant in each phase. 
We used an on-line tool developed by 
\cite{Robitaille+etal+2007} for SED modeling. Using this tool, several authors have successfully modeled  
T Tauri stars and massive protostars with masses up to 25 M$_{\odot}$ (e.g., 
\cite{Dewangan+Anandarao+2010}). The on-line tool 
selects the best-fit solutions from 20,000 models (each with 10 different angles of inclination for the accretion disk, 
making a total of 200,000 models). 
The input parameters include, apart from a minimum of three data points in SED 
and their corresponding errors, range of distances to the object and the visual 
extinction. The output parameters include stellar mass, temperature, radius, age and the total 
luminosity; as well as disk inclination angle, the disk and envelope masses and rates of accretion. 
In order to minimize the degeneracy 
of solutions, only those solutions that satisfy the criterion 
\begin{equation}
(\chi^{2} - \chi^{2}_{best}) \leq 3 
\end{equation}
\noindent are chosen with $\chi^{2}$ considered per data point (see  \cite{Robitaille+etal+2006}). 
Further, in order to avoid `over-interpretation' of 
SEDs, we provided a small range of visual extinction values (from the JHK photometry) 
for each phase, to account for the inherent uncertainties in their determination. In the present case, 
a distance of 0.40 kpc is adopted (from \cite{Anthony-Twarog+1982}). 
The best fit models are displayed in Fig 7 along with the observed data (shown in filled 
circles) for the quiescent phase of 2006-2007 and the on-going outburst phase of 2008-2012.
 
Table 2 lists the mean values and standard deviations of 
physical parameters derived from the models for the two epochs. 
The model parameters for the outburst phase 2004-2005 match quite well with those of the 
on-going outburst phase of 2008-2012.
In the quiescent phase one can notice the stellar parameters similar to those of a 
T Tauri star. But the outburst parameters mimic more of an intermediate mass star rather than a low mass 
PMS star. In the outburst phase the disk mass and accretion rate are enhanced substantially
when compared to the quiescent phase. The envelope accretion rate decreases by more than 
an order of magnitude in the outburst phase compared to the quiescence while
envelope mass decreases by several orders of magnitude. This may be due to 
the enhanced luminosity in the outburst phase which could prevent mass accretion from 
the envelope. In the quiescent phase, the envelope emission dominates in the mid-infrared 
and longer wavelength region; while in the outburst phase the disk emission contributes 
substantially in the region beyond 1 $\mu$m.

It may be noted that the disk accretion rate for the outburst phase given in Table 2 
matches well with those reported by \cite{Briceno+etal+2004}, \cite{Aspin+etal+2008}, \cite{Brittain+etal+2010} and 
\citet{Aspin+2011}. However, the disk accretion rates
in the quiescent phase derived from Br$\gamma$/Pa$\beta$ emission lines by \cite{Aspin+etal+2008} 
are much higher than the model-derived value reported here. While this could be due to uncertainties in the
SED model or emission line method, the important point to be noted is 
that the disk accretion rate during the outburst phase is much larger than that in the quiescent period. 
The total luminosity obtained here (which includes the contributions from the star and its disk and envelope) 
matches well with those derived by \cite{Abraham+etal+2004} and \cite{Aspin+etal+2008} from the SEDs of the pre-outburst
phase.   
The total outburst luminosity during the 2003 outburst as derived by \cite{Briceno+etal+2004} from 
the SED is 219 L$_{\odot}$ which agrees well with that obtained from our model for the current on-going outburst.
Earlier, using a simpler model, \cite{Muzerolle+etal+2005} interpreted the
SPITZER IRAC/MIPS (photometric bands between 3.5-70 $\mu$m) data for the pre- and post-outburst phases. 
Their results are qualitatively similar to ours.

\section{Conclusions}
\label{sec:concl}
The important conclusions of this work are: 
\begin{enumerate}
\item{Monitoring of V1647 Ori in the {\it JHK} bands has shown that the object has been undergoing 
episodes of mass accretion variation as indicated by small but significant variations 
in its {\it JHK} fluxes. This conclusion is also supported by its near-infrared spectral variations;}
\item{Starting from early 2012, the object seems to show a slow fading out with a rate of 
$\sim$ 0.06-0.18 mag per year. This may indicate that V1647 Ori is an 
intermediate type object falling between FUors and EXors, having characteristic features of 
both the proto-types;}
\item{{\it JHK} color-color diagram indicates several occasions in the current, on-going outburst 
in which the star displays positions beyond the T Tauri region indicating the presence of colder dust
compared to its 2004 outburst phase;}   
\item{The Herbig-Haro object HH22, whose as yet unidentified energizing YSO is  
situated at about 30 arcsec from V1647 Ori seems to show light 
fluctuations corresponding to those of the latter, thereby confirming that it is a reflection 
nebulosity triggered by the YSO;} 
\item{The spectral energy distributions of V1647 Ori modeled for the epochs of 
quiescence and outburst, show that the disk mass and accretion rate in the outburst 
phase are significantly greater than that in the quiescent stage, while the envelope mass and 
accretion rates are much lower.}
\end{enumerate} 

\normalem
\section*{acknowledgments}
The research work at PRL is supported by the Department of Space, Government of 
India. It is a pleasure to thank the Mt Abu Observatory staff for their cooperation 
in making observations. We gratefully acknowledge using the SEDFIT online software 
developed by Robitaille et al (2007). We thank the anonymous reviewer for useful comments.

\bibliographystyle{raa}
\bibliography{v1647ori_2}

\end{document}